\documentclass[superscriptaddress, twocolumn, amsmath, amssymb, aps, prl, letterdraft, notitlepage, floatfix
]{revtex4-2}
\usepackage{braket}
\usepackage{graphicx}% Include figure files
\usepackage{dcolumn}% Align table columns on decimal point
\usepackage{bm}% bold math
\usepackage{hyperref}% add hypertext capabilities
\usepackage{newfloat} % must be loaded before algorithm
\DeclareFloatingEnvironment[
    fileext=loa,           % extension for list-of-algorithm file
    name=Algorithm         % caption name
]{algorithm}
\usepackage[main=english]{babel}
\makeatletter
\ifcsname l@english\endcsname
  \expandafter\let\csname l@eng\endcsname\l@english
\fi

\makeatother
\usepackage[utf8]{inputenc}
\usepackage{algpseudocode}
\usepackage{multirow}
\usepackage{diagbox}
\usepackage{ulem}
\usepackage{xcolor}
\usepackage{natbib}

\begin{document}

\preprint{APS/123-QED}

\title{Neural Network Assisted Fermionic Compression Encoding: A Lossy-QSCI Framework for Resource-Efficient Ground-State Simulations}% Force line breaks with \\

\author{Yu-Cheng Chen}
\email{kesson.yc.chen@foxconn.com}
\affiliation{Hon Hai Research Institute, Taipei, Taiwan}

\author{Ronin Wu}%
\email{ronin@qunasys.com}
\affiliation{%
QunaSys Europe, Copenhagen, Denmark
}%

\author{M. H. Cheng}
\affiliation{Department of Physics Blackett Laboratory, Imperial College London, SW7 2AZ, United Kingdom}
\affiliation{Fraunhofer Institute for Industrial Mathematics, Fraunhofer-Platz 1, 67663 Kaiserslautern, Germany}

\author{Min-Hsiu Hsieh}
\affiliation{Hon Hai Research Institute, Taipei, Taiwan}

% Man Hei Cheng's commands
\newcommand{\cheng}[1]{\textcolor{blue}{#1}}
\newcommand{\lossyqsci}{\textsc{Lossy-QSCI}}
\newcommand{\chemrle}{\textsc{Chemical-RLE}}
\newcommand{\nnfed}{\textsc{NN-FED}}
\newcommand{\onlogm}{\( O(N \log M) \)}
\date{\today}% It is always \today, today,

\begin{abstract}
Quantum computing promises to revolutionize many-body simulations for quantum chemistry, but its potential is constrained by limited qubits and noise in current devices. In this work, we introduce the Lossy Quantum Selected Configuration Interaction (\lossyqsci{}) framework, which combines a lossy subspace Hamiltonian preparation pipeline with a generic QSCI selection process. This framework integrates a chemistry-inspired lossy Random Linear Encoder (\chemrle{}) with a neural network-assisted Fermionic Expectation Decoder (\nnfed{}). The RLE leverages fermionic number conservation to compress quantum states, reducing qubit requirements to \onlogm{} for \( M \) spin orbitals and \( N \) electrons, while preserving crucial ground state information and enabling self-consistent configuration recovery. \nnfed{}, powered by a neural network trained with minimal data, efficiently decodes these compressed states, overcoming the measurement challenges common in the approaches of the traditional QSCI and its variants. Through iterative quantum sampling and classical post-processing, our hybrid method refines ground state estimates with high efficiency. {This framework offers a resource-efficient pathway for ground-state simulations on near-term noisy hardware and could inspire resource-efficient extensions to future devices by minimizing qubit overhead.}
\end{abstract}

\maketitle

%\tableofcontents

\section{Introduction}

Quantum computing enables precise emulation of large-scale electronic structures problems, delivering promises to advance quantum chemistry\cite{aspuru-guzik_simulated_2005}. This potential stems from the natural compatibility of quantum computers in representing quantum many-body systems, a task intractable for classical computers in many cases\cite{feynman_simulating_1982, lloyd_universal_1996}. However, in the presence of quantum noise,  near-term quantum devices face tremendous obstacles in the quest to realize this promise. Hybrid algorithms such as the Variational Quantum Algorithm (VQA)\cite{peruzzo_variational_2014, tilly_variational_2022} and Quantum Selected Configuration Interaction (QSCI)\cite{kanno_quantum-selected_2023} subsequently emerged as counter-strategies, harnessing the power of quantum sampling to approximate molecular ground states. 

VQA, for example, exploits parameterized quantum circuits\cite{du2020expressive} combined with classical optimization to converge to the ground state energy iteratively. Recent advances have demonstrated improved ansatz design\cite{kanasugi_computation_2023, mizuta_local_2022, li2023gell, du2022quantum}, enhanced error mitigation\cite{inoue_almost_2023}, and reduced measurement overhead\cite{heya_subspace_2023}, thereby bolstering VQA’s robustness against noise and its scalability for simulating complex electronic correlations. These developments reinforce the promise of hybrid algorithms in overcoming hardware limitations and paving the way for more efficient quantum chemistry simulations. Although VQA shows promise, it faces serious challenges. Its variational optimization is highly sensitive to quantum noise, prone to barren plateaus, and saddled with a measurement overhead that grows rapidly with system size. These limitations call for alternatives. 

In contrast, QSCI leverages quantum sampling to construct effective subspaces for diagonalization and uses chemical priors to enhance state selection and energy estimation. It thereby addresses the noise sensitivity and high measurement cost issues inherent in Variational Quantum Eigensolver (VQE), and offers a new pathway for accurate electronic structure simulations. However, traditional QSCI does not compress the qubit space; it requires the full Hilbert space representation, leading to high resource demands and substantial classical post‐processing overhead that remains sensitive to noise. To overcome this, it is demonstrated in a scaled-up QSCI experiment (77 qubits) that by incorporating self‐consistent configuration recovery and employing chemical priors via the localized unitary cluster Jastrow (LUCJ) ansatz\cite{motta_bridging_2023}, one can improve noise robustness; this yields higher state quality but adds complexity and further strains classical resources\cite{robledo-moreno_chemistry_2024}. 

On the other hand, ADAPT-QSCI~\cite{nakagawa_adaptqsci_2024} iteratively grows the input state by ADAPT-VQE‐style operator selection, thereby removing the need for a fixed chemical ansatz but at the cost of many optimization rounds and repeated QSCI executions. More recently, Time‐Evolved QSCI (TE-QSCI)\cite{mikkelsen_quantum-selected_2025} and concurrent work by \citet{sugisaki_hamiltonian_2025} have been proposed to prepare the QSCI input state via time evolution instead of variational optimization, naturally generating a compact subspace rich in electronically excited configurations. Later, \citet{yu_quantum-centric_2025} independently adopted the same “time-evolved input-state’’ motif, signifying growing interest in this direction. By reducing measurement overhead through effective selections of important configurations, TE-QSCI offers a more resource-efficient alternative that complements and, in some aspects, outperforms the scaled-up QSCI presented in \citet{robledo-moreno_chemistry_2024} in balancing qubit efficiency and overall computational cost. 

Nonetheless, even with these advancements, QSCI and its variants still suffer from sampling inefficiency. The full Hilbert space representation results in less compact wavefunctions and higher qubit resource demands\cite{reinholdt_fundamental_2025}. Moreover, the inherent complex ansatz design and the variational optimization in standard QSCI or even TE-QSCI introduces additional overheads and challenges that strain both classical and quantum resources. These issues strongly motivate our work in advanced fermionic encoding methods, which aim to directly compress the qubit space while preserving the essential physics of the electronic Hamiltonian. By addressing both the qubit resource inefficiency and the burden of complex ansatz design, our approach offers a more resource-efficient pathway for ground state simulations.

Traditional fermionic encodings, such as the Jordan-Wigner (JW) transformation \cite{jordan_ber_1928}, capture the full Fock space of electronic Hamiltonians, resulting in linear qubit scaling with system size. Recent number-conserving encodings \cite{shee_qubit-efficient_2022} perform a log-qubit scaling by restricting the simulation to the number-conserving subspace or even compact Configuration Interaction (CI) subspace~\cite{yoffe2024qubit}. Specifically, focusing on states with a fixed electron number $N$ from the $M$ available spin orbitals, the effective number of qubits needed is roughly given by $\log_2 \binom{M}{N}$. For cases where $N\ll M$, this quantity scales approximately as the \onlogm{} scaling. However, they incur an increased energy measurement overhead, typically scaling as the number of states(e.g., \( O(N ^ M) \) for number-conserving), due to the need to project onto these conserved subspaces, which restricts the scaling efficiency of solving the subspace Hamiltonian. The Fermionic Expectation Decoder (FED) has offered a partial solution by decoupling measurement scaling from electron number theoretically but has yet to be realized through implementation~\cite{cheng_optimal_2024}. A related first-quantized approach also attains the nominal \onlogm{} qubit scaling by encoding an antisymmetrised $N$-electron wave-function in a register of $N\log_2 M$ qubits\cite{babbush2017exponentially}. However, because the wave-function lives in first quantization, Hamiltonian application and the method are not directly compatible with selected-CI workflows. These limitations highlight the need for innovative approaches that optimize both qubit usage and measurement efficiency.

Motivated by these limitations, we present a \lossyqsci{} framework that integrates quantum encoding and classical decoding techniques. This approach compresses the qubit space while mitigating measurement overhead and preserving essential chemical information. Central to this approach is a new chemistry-inspired lossy Randomized Linear Encoding (\chemrle{}) modified from the original lossless RLE \cite{cheng_optimal_2024}, which lossy compresses quantum states while preserving essential chemical properties, substantially reducing qubit cost and scaling. In complement, a neural network-assisted FED (\nnfed{}) is introduced to rapidly and accurately decode compressed states using minimal training data, surpassing traditional classical decoding methods. The framework operates iteratively, sampling compressed quantum states and refining ground state estimates through classical post-processing. Our method achieves chemical accuracy with fewer resources on \( \mathrm{C}_2 \) and \( \mathrm{LiH} \) molecules, demonstrating its efficacy and accuracy.

Table~\ref{tab:comparison} presents a comparative analysis of the original QSCI, its variants, and the Lossy-QSCI framework. For qubit efficiency, \lossyqsci{} excels in reducing the qubit needs to \onlogm{} or fewer via \chemrle{}, while others require the M qubits. For error mitigation, all approaches address bit-flip errors that violate the conservation of particle number. QSCI and TE-QSCI use post-selection to reduce errors, and the scaled-up QSCI employs self-consistent configuration recovery to further mitigate them, whereas \lossyqsci{} leverages qubit compression to preclude the inclusion of non-conserving configurations. Incorporating chemical priors, QSCI and its scaled-up version utilize a chemical ansatz to constrain the solution subspace and boost sampling efficiency, while TE-QSCI relies on a carefully chosen ansatz (e.g., Hartree-Fock or UCCSD) combined with time evolution to generate a configuration-rich subspace naturally. On the other hand, \lossyqsci{} relies on its chemistry-informed encoding strategy, \chemrle{}.

{In practice, the advantages of \lossyqsci{} rely on two empirical steps: a biased, chemistry-guided injectivity check and a neural decoder trained on a limited set of sample determinants. Both procedures scale favorably for the benchmarked medium-sized systems studied here, but they are not yet universal tools that guarantee success for arbitrary, much larger molecules. Consequently, the present framework should be viewed as a pragmatic extension of existing hybrid schemes (VQE, QSCI, etc.), whose own demonstrations have so far been confined to $\lesssim10^{2}$ physical qubits and well-characterized Hamiltonians. Within these limitations, \lossyqsci{} still delivers a worthwhile trade-off between qubit count, measurement overhead, and final accuracy. It provides the first proof-of-concept algorithm that demonstrates improved configuration sampling not only via ansatz choice, but also encoding choice. Such a classical-quantum interface not only improves the expressivity of variational algorithms by using the classical encoding degrees of freedom, but it also significantly reduces the hardware burdens. By identifying encoding as a quantum resource for chemistry computation, we extend the potential quantum advantage for variational quantum algorithms.}

\begin{table*}[t]
\scriptsize
\centering
\begin{tabular}{|c|c|c|c|c|}
\hline
\textbf{Method} & \textbf{QSCI~\cite{kanno_quantum-selected_2023}} & \textbf{Scaled-up QSCI\cite{robledo-moreno_chemistry_2024}} & \textbf{TE-QSCI\cite{mikkelsen_quantum-selected_2025}} & \textbf{\lossyqsci{} [This work]} \\ \hline
Qubit Efficiency & O(M) & O(M) & O(M) & O(Nlog(M)) \\ \hline
Bit-flip Error Mitigation & Post-Selection & Configuration Recovery & Post-Selection & Compression+Post-Selection \\ \hline
Chemical Prior & Customizable & LUCJ Ansatz & Time-Evolved-HF/UCCSD & \chemrle{} \\ \hline
\end{tabular}
\caption{
Comparison of Original QSCI, its variants, and \lossyqsci{} Frameworks.
Criteria are defined as follows: 
\textbf{Qubit Efficiency}: Denotes optimization of qubit usage. 
\textbf{Bit-flip Error Mitigation}: How the method mitigates bit-flip errors with particle number conservation. 
\textbf{Chemical Prior}: Specifies how prior chemical knowledge constrains the solution subspace. 
}
\label{tab:comparison}
\end{table*}

\section{Previous works}
\subsection{Fermionic Encoding}
{Let us consider a second quantized quantum chemistry Hamiltonian, comprising the fermionic creation and annihilation operators \( \hat{a}_{p\sigma}^\dagger \) and \( \hat{a}_{p\sigma} \), and the one and two electronic integrals \(h_{pr}, h_{pqrs}\) associated\citep{szabo_modern_1996}: 
\begin{equation}
\hat{H} = \sum_{p,r,\sigma} h_{pr} \hat{a}_{p\sigma}^\dagger \hat{a}_{r\sigma} + \frac{1}{2} \sum_{\substack{p,r,q,s \\ \sigma,\tau}} h_{pqrs} \hat{a}_{p\sigma}^\dagger \hat{a}_{q\tau}^\dagger \hat{a}_{s\tau} \hat{a}_{r\sigma},
\label{eqn: hamiltonian}
\end{equation}
where $h_{pq}$ and $h_{pqrs}$ are one- and two-electron integrals over spin orbitals (indices p, q, r, s), regarding to quantum chemistry standards; $\sigma$ denotes spin indices (up or down). The Hamiltonian finds a natural encoding onto the qubit system via JW transformation, resulting in a linear combination of unitary $\hat{H} = \sum_{i}c_{i}U_i$ for the expectation estimation \cite{jordan_ber_1928}, where the index i refers to the i-th unitary operator in the ansatz decomposition.} Given the system with $M$ spin orbitals and $N$ electrons, it has been shown that the number-conserving encoding reduces the qubit resource requirements up to \onlogm{} scaling in both linear and non-linear compression schemes \cite{shee_qubit-efficient_2022, cheng_optimal_2024}. These schemes can be collectively denoted as an encoder $\mathcal{E}(\cdot)$ that maps the JW basis $\{\vec{b_i} \}\in \{0, 1\}^M$ of $\hat{H}$ to $\{\mathcal{E}(\vec{b_i})\}$ that preserves the N electron number-conserving basis states $\mathbf{b_N}$. Linear encoding further preserves the linear bitwise addition property to crucial for describing fermionic transitions. $\mathcal{E}(b_i \oplus b_j)=\mathcal{E}(b_i)\oplus\mathcal{E}(b_j).$ Although qubit-efficient type non-linear compression achieves optimal qubit compression rate \cite{shee_qubit-efficient_2022}, the number of measurement bases will scale up to $O(M^N)$.

With linear encoding, operations can be avoided by performing all computations in the compressed subspace. The existence of an efficient decoder guarantee that map-encoded quantum states $\mathcal{E}(b_i)$ back to $b_i$ at least for all number-conserving states $b_i\in\mathbf{b_N}$, {where $\mathbf{b_N}$ is a bitstring representation for N-electron configurations.} That is, 
\begin{equation}
    \mathcal{D}(\mathcal{E}(s_i))=s_i.
    \label{eqn: decode-encode}
\end{equation}
This observation leads to a new expectation estimation method -- the FED algorithm \cite{cheng_optimal_2024}. This method also guarantees at most $O(M^4)$ measurement bases required for sampling each element of fermionic reduced density matrices (RDMs). These expectation values can be measured, classically post-processed, and computed in polynomial time. Meanwhile, operation in the compressed subspace reduces the impact of quantum noise since (i) number-conserved bitstrings are more likely mapped to number-conserved bitstrings, and (ii) the qubit cost significantly reduces. 

However, there are still two key challenges for the optimal linear encoding framework. First, encoding requires N-electron states to be injectively mapped to the encoded states. Utilizing random linear code, encoder generation will require the worst-case $O(M^N)$ checks. Although there are deterministic decoding methods, they either require access to a sparse oracle \cite{harrow_quantum_2009} or require a great number of qubits (Polylog) \cite{childs_quantum_2017} to achieve an advantage. Second, decoding this problem is generally hard, the naive way is to store the look-up table to get optimal time complexity while trading off with space complexity $O(M^N)$ to restore the data. While there exist efficient methods widely used in quantum error correction (QEC) \cite{shor_scheme_1995, steane_error_1996, gottesman_stabilizer_1997}, they are not specifically designed for this problem, since we only want to ensure the decoding of $\mathbf{b_N}$ is correct. 

\subsection{QSCI Framework}

QSCI algorithms—including the 77-qubit demonstration and the more recent TE-QSCI—show how near-term hardware can tackle quantum chemistry. They sample a trial state $\ket{\Psi}$ on a quantum device, retain the most probable bitstrings ${\mathbf{x}_i}$, and use these determinants to define a reduced configuration–interaction subspace in which the electronic Hamiltonian $\hat{H}$ is classically diagonalised to obtain ground-state energies.

To address the sensitivity of QSCI to the \emph{a-priori} choice of trial state (unclear), an \emph{adaptive} variant has been introduced. ADAPT-QSCI~\cite{nakagawa_adaptqsci_2024} mirrors the ADAPT-VQE strategy. At each macro-iteration it ranks all single- and double-excitation operators by the size of their QSCI energy gradient, appends the best candidate to the circuit, then runs a brief classical re-optimization before moving to the next iteration. Because the ansatz is grown on-the-fly, ADAPT-QSCI alleviates the “good initial state” problem, but this comes at the cost of many repeated QSCI and gradient evaluations, and no saving in qubit count since the full JW register is still required. Beside the adaptive approaches, LUCJ ansatz has also been proposed to tackle hardware noise by concentrating probability of \( |\Psi\rangle \) in configurations near the ground-state support, enabling efficient sampling even on noisy devices~\cite{robledo-moreno_chemistry_2024}.

Beyond the variational approaches, TE-QSCI represents an alternative to initial state preparation problems. Time evolution operator $e^{-i\hat{H}t}$ is employed to an initial state, such as Hartree–Fock or UCCSD, naturally generating a subspace rich in electronically excited configurations. This method avoids overheads from iterative optimization and the recurring use of quantum circuits that plagues the variational methods, while ensuring sampling in the physical effective subspace necessary for QSCI diagonalization. 

Overall, the computational cost of all QSCI variants critically depends on the overlap between trial state \( |\Psi\rangle \) with the ground-state wavefunction. Let \( P_G \) represent the Ground-State Projector, which indicates the probability of the operator projecting onto the low-energy subspace. If \( |\Psi\rangle \) preferentially samples configurations \( P_G \), the resulting subspace diagonalization yields accurate energy estimates with samples that scale polynomially with system size. 

Despite notable progress, all existing QSCI variants exhibit two fundamental limitations. First, they are qubit-inefficient: current methods encode the full Fock space using one qubit per fermionic mode, rapidly depleting quantum hardware resources as the system size increases. Second, they depend on complex ansatz constructions or iterative optimization routines—such as the multi-term LUCJ ansatz or repeated state preparation in TE-QSCI—which lead to high circuit depth and significant classical overhead. These challenges motivate our \lossyqsci{} approach. By leveraging \chemrle{}, we compress the qubit register to \onlogm{} and employ a lightweight \nnfed{} decoder to recover and measure only the most relevant configurations. This direct compression in qubit space, coupled with efficient decoding, addresses both the qubit-scaling and ansatz-complexity bottlenecks of conventional QSCI frameworks.

\begin{figure}
    \centering
    \includegraphics[width=1.0\linewidth]{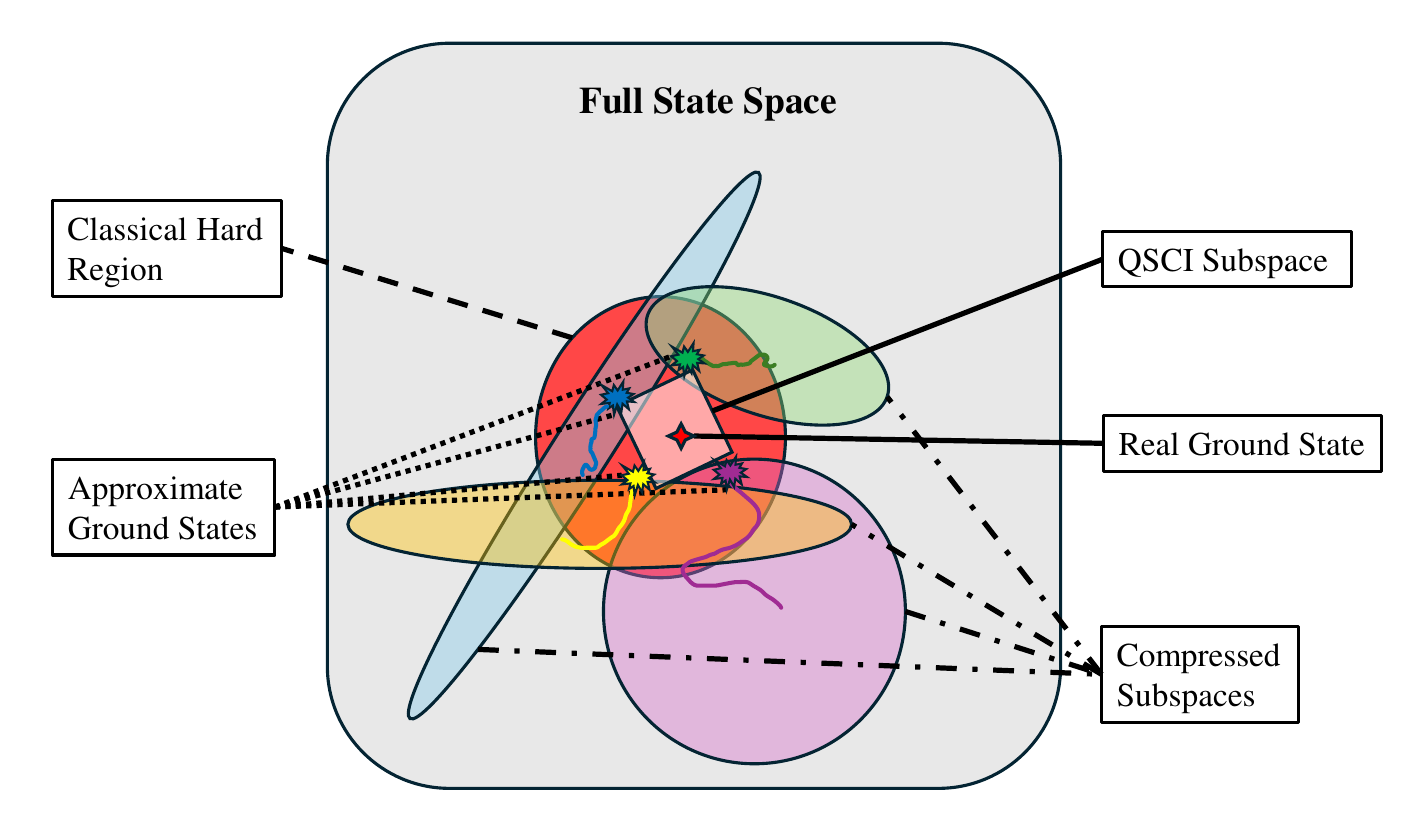}
    \caption{This diagram illustrates the key concept of \lossyqsci{}. RLE is preformed around the ground state to perform VQE. The resulting minima are sampled to gauge the region of the QSCI subspace.}
    \label{fig1}
\end{figure}
\section{\lossyqsci{}}
In this work, we introduce the \lossyqsci{} framework. It enhances quantum chemistry simulations by integrating a chemistry-inspired \chemrle{} with a neural network FED, building on the strengths of fermionic encodings. Leveraging the number-conserving basis \( \mathbf{b_N} \) from prior sections—where qubit requirements scale as \onlogm{} for \( M \) spin orbitals and \( N \) electrons— RLE compresses basis states \( \{\vec{b_i}\} \) into a qubit-efficient representation \( \{\mathcal{E}(\vec{b_i})\} \), preserving a bijective mapping on important chemical subspace and avoiding the \( O(M^N) \) measurement overhead of nonlinear encoding. To decode these compressed states efficiently, we propose a NN decoder trained on the structured patterns of electronic configurations, sidestepping the combinatorial space complexity of naive methods and improving the performance of the FED. The use of \chemrle{} naturally integrates QSCI with particle-conserving post-selection by using the Q-qubit compressed states as trial wavefunctions \( |\Psi\rangle \in F_2^Q \), sampling configurations to build subspaces for diagonalization. It can further enhance accuracy by adopting a lossy subspace as demonstrated in Fig. \ref{fig1}, concentrating probability on configurations near the ground state with even fewer qubits, ensuring significant overlap with the true solution, and enabling precise energy estimation with polynomially many samples.

\subsection{Chemistry-Inspired \chemrle{}}
To integrate bit-flip error mitigation and \textit{ground state concentration} in QSCI with RLE, we provide a new \chemrle{} design that can work with QSCI to approach the full space ground state by sampling the subspace approximate ground state to improve usability. The original RLE begins its random search by choosing a $Q$ within the bounded region $N\log_2 M < Q < 2N\log_2 M$. Then, the algorithm initializes the parity check matrix $\mathbf{G} = [I_Q | D]$ in the standard basis, where $I_Q$ is a $Q \times Q$ identity matrix. In previous work\citep{cheng_unleashing_2023}, the $Q \times (M-Q)$ matrix $D$ is randomly generated such that each column has an even Hamming weight equal to the value $Q/2$, denoted as $\text{even}(Q/2)$. However, such rules restrict the compressibility of qubit numbers, and the original RLE also requires a heavy subroutine check to ensure lossless configurations during compression. To address this, we propose a lossy compression scheme without a bijective checking process or with a biased bijective checking, ensuring the compressed Hamiltonian retains sufficient information. This approach is particularly well-suited for quantum chemistry, where ground states are often sparse in a well-defined basis and conform to specific symmetries or chemical rules(e.g., classical selected CI), making them ideal for lossy compression with minimal loss of critical information.

For the bit-flip error mitigation, encoding particle-number symmetry at the RLE stage removes much of the Hilbert-space redundancy that the original QSCI exploited for post-selection. In the uncompressed mapping, a single bit-flip typically drives the register outside the selected $N$-electron sector, so post-selection suppresses logical errors from $O(p)$ to $O(p^{2})$ (see App.~B of Ref.~\cite{kanno_quantum-selected_2023}).  After compression, however, a single bit-flip can map one valid encoded determinant to another, restoring an $O(p)$ error rate.  We compensate for this loss by: (i) discarding any decoded string that falls outside the Complete Active Space(CAS)-biased domain ignored by the lossy map, and (ii) rejecting shots whose decoder likelihood or estimated energy lies beyond predefined thresholds.

For the preparation of a compact wavefunction with chemical knowledge, one of the selections is to use a classical heuristic\cite{tubman2020modern} for chemical-inspired lossy RLE. By biased bijective checking, RLE will be enforced to generate a compressed Hamiltonian span by the classical heuristic. One can then further use QSCI to filter out more configurations for preparing even compact CI expansions. Another design is backed by the Molecular Orbitals~(MOs) Theory\cite{szabo_modern_1996}. Since the chemical Hamiltonian's ground state normally has a biased distribution, configurations with lower energy molecular orbitals will contribute most to the ground state wavefunction. With this, the CAS method~\cite{roos_complete_1980} has been introduced to remove the high-energy MOs for better resource usage. Inspired by this fact, we can easily construct the \chemrle{} that preserves the lower energy configurations while randomly losing the configurations that might have higher energy. By setting the computing basis with MOs order from lowest energy to highest energy as $$\ket{1_{u}1_{d}2_{u}2_{d}...M_{u}M_{d}},$$ where number is energy ordered MOs, u is spin up orbitals and d is spin down orbitals. Combined with the original RLE design that can prepare a lossless map of N-electron configurations whose electrons are all in the front Q orbitals as $$\ket{1_{u}1_{d}2_{u}2_{d}...Q_{u}Q_{d}000...000},$$ the $I_Q$ part in $\mathbf{G}$ can guarantee the lossless while electron belongs in D part. Then, we can further reduce the checking process by bias sampling the one-electron($\{\cdot\}_1$), two-electron($\{\cdot\}_2$), ..., N-electron excited state with different probabilities,
\begin{align*}
     \{\ket{1_{u}...Q_{d}100...000},&...,\ket{1_{u}...Q_{d}000...001}\}_1 \\
    \{\ket{1_{u}...Q_{d}110...000},&...,\ket{1_{u}...Q_{d}000...011}\}_2 \\
    &\vdots \\
    \{\ket{1_{u}...Q_{d}111...100},&...,\ket{1_{u}...Q_{d}001...111}\}_N
\end{align*}

This basis selection provides two advantages if the ground state is \textit{sparse}. First, it can further reduce the number of qubits lower than the theoretical low bounds $log_2(\binom{M}{N})$ while still being able to recover the ground state support. Second, it can skip the costly injectivity checking process by not checking or only checking the important states from the predefined prior subspace. 

\begin{figure}
    \centering
    \hspace*{-0.5cm}\includegraphics[width=0.95\linewidth]{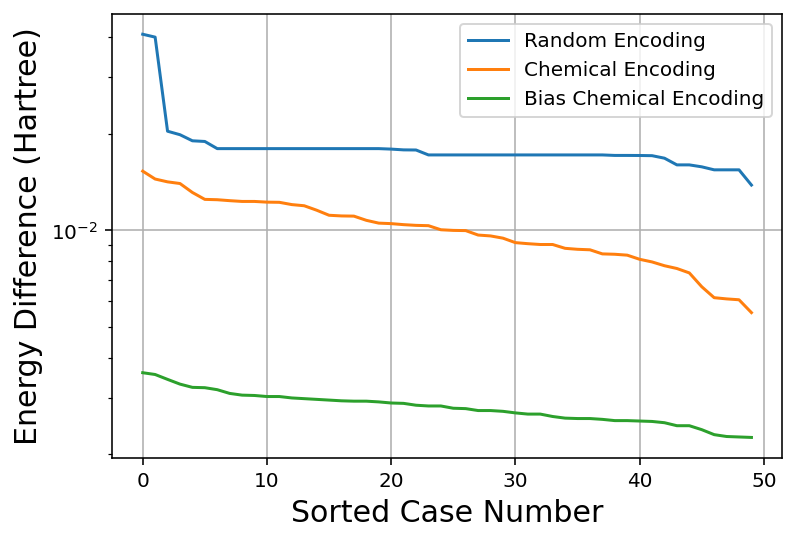}
    \caption{Result of different RLE strategies performance. The X-axis represents the case number sorted by the final converged energy of QSCI, and the Y-axis denotes the energy difference from the exact ground state.}
    \label{fig2}
\end{figure}

To test the performance of \chemrle{}, we compared three different strategies for solving 16 Spin orbitals and 6 electrons $C_2H_4$ system with 12 qubits ($log_2(\binom{16}{6})$), see Fig. \ref{fig2}. The first strategy, Random Encoding, involves randomly selecting the computing basis without considering the energy-sorted order of MOs and without any injectivity checking. The second strategy, Chemical Encoding, uses the energy-sorted basis order of MOs without injectivity checking. {The third strategy, Biased Chemical Encoding, employs the energy-sorted basis order of MOs, incorporating pre-selected configurations which are considered more likely to be ground state support for injectivity checking. In this test, we select the bias injectivity checking states from the top 200 most frequent configurations of ground states. In a real case simulation scenario, pre-selected configurations could be obtained from prior knowledge, e.g., derived from classical approximations like Hartree-Fock, Selected Configuration Interaction, or Complete Active Space Self-Consistent Field (CASSCF) methods, which identify chemically relevant subspaces and symmetries of the system. A random uniform noise is added to the exact ground state to prepare an approximate ground state with about 0.9 fidelity. This process is repeated 50 times for \lossyqsci{}, each performed 20 times, and the results are collected to prepare the effective Hamiltonian for classical diagonalization. This result shows that the RLE that guarantees bias injectivity could provide better lossy compression quality. It also shows that more prior knowledge of the given chemical system could lead to a better lossy compression rate since the bias injectivity can be more constrained to a smaller subspace.}

\subsection{Efficient Neural Network Decoder, \nnfed{}}
\begin{table*}[]
\begin{tabular}{|c|c|c|c|c|c|c|}
\hline
\diagbox[dir=SE]{Algorithms}{4-Electron}             & \textbf{(M, Q)}                  & \textbf{(30,23)} & \textbf{(40,27)} & \textbf{(50,30)} & \textbf{(60,32)} & \textbf{(70,34)} \\ \hline
\multirow{3}{*}{\nnfed{}}          & \multicolumn{1}{c|}{Training Time (min)} & 2.56    & 5.09    & 6.97    & 11.50   & 14.22   \\ \cline{2-7} 
                                     & Decoding Times (sec)                     & 0.0022  & 0.0022  & 0.0032  & 0.0076  & 0.0072  \\ \cline{2-7} 
                                     & Accuracy                                 & 100.0\% & 99.9\%  & 100.0\% & 99.8\%  & 99.3\%  \\ \hline
\multirow{2}{*}{Genetic Algorithm}   & Decoding Times (sec)                     & 3.6693  & 3.9568  & 3.6679  & 3.6423  & 3.7163  \\ \cline{2-7} 
                                     & Accuracy                                 & 90.4\%  & 83.9\%  & 76.0\%  & 71.8\%  & 65.0\%  \\ \hline
\multirow{2}{*}{Simulated Annealing} & Decoding Times (sec)                     & 0.0845  & 0.1017  & 0.1112  & 0.1187  & 0.1206  \\ \cline{2-7} 
                                     & Accuracy                                 & 40.3\%  & 31.5\%  & 24.3\%  & 18.3\%  & 16.1\%  \\ \hline
\end{tabular}
\caption{This figure illustrates the \nnfed{} benchmarks. M is the number of spin orbitals, and Q is the number of qubits.}
\label{t1}
\end{table*}

Beyond the injectivity check required by \textsc{RLE}, the main bottleneck in \lossyqsci{} is the decoding stage of the \textsc{FED}.  Reconstructing an \(M\)-qubit occupation string from its \(Q\)-qubit codeword is difficult because the number of admissible states grows as \(\mathcal{O}(M^{N})\).  A naïve lookup table demands exponential memory, whereas heuristic optimizers (e.\,g., simulated annealing or genetic algorithms) yield only approximate solutions and scale poorly.

To provide an efficient alternative, we introduce a neural-network–assisted decoder (\nnfed{}), trained with the workflow in Algorithm~\ref{alg1}.  We first fix a random linear encoder \(\mathcal{E}(\cdot)\) and initialize a feed-forward network \(\mathcal{D}_{\vec{\theta}}(\cdot)\) that maps a \(Q\)-qubit codeword to an \(M\)-qubit bitstring; the network contains \(\mathcal{O}(MN)\) parameters in all test cases.  During each training iteration we draw \(k\) random \(N\)-electron strings \(\{s_i\}\), encode them, and let the network predict \(\{\hat{s}_i\}\).  The parameters \(\vec{\theta}\) are updated by minimising the binary cross-entropy (BCE) \citep{bishop_pattern_2006}
\[
L_{\mathrm{BCE}}
  =-\frac{1}{k}\sum_{i=1}^{k}
    \bigl[
      s_i\cdot\log\hat{s}_i
      +(1-s_i)\cdot\log\!\bigl(1-\hat{s}_i\bigr)
    \bigr].
\]
After a few epochs, the converged network \(\mathcal{D}_{\vec{\theta}}\) becomes the fast decoder \(\mathcal{D}_{\mathrm{NN}}\) employed by the \textsc{FED} in Eq.~\eqref{eqn: decode-encode}.

\begin{algorithm}
    \caption{\nnfed{}}
    \label{alg1}
    \begin{algorithmic}
    \Require
       Random Linear Encoder $\mathcal{E}(\cdot)$, Random N-electron M-qubit bit string generator $\mathbf{S}^M_N\in F^M_2$;
    \Ensure
        $\mathcal{D_{NN}}(\mathcal{E}(s_i))=s_i, \forall s_i \in \mathbf{S}^M_N$;
    \State Initialize a Neural Network 
    \State $\mathcal{D_{\vec{\theta}}(\cdot)}:F^Q_2 \to  F^M_2$;
    \For{ $t < t_{max}$}    
        \State
        Randomly sample 
        \State
        $\{s_i^{(t)}\}\in\mathbf{S}^M_N$;
        \State
        Prepare training data 
        \State
        $\{b_i^{(t)}\} \leftarrow  \mathcal{E}(\{s_i^{(t)}\}) \in F^Q_2$;
        \State
        $\{\hat{s}_i^{(t)}\} \leftarrow \mathcal{D_{\vec{\theta}}}(b_i^{(t)}) \in F^M_2$;
        \State
        $ \Delta \vec{\theta} \leftarrow \nabla_{\vec{\theta}^{(t)}}L_{BCE}(\{s_i\},\{\hat{s}_i\})$
        \State
        $\vec{\theta}^{(t+1)}\leftarrow \vec{\theta}^{(t)} + \Delta \vec{\theta}$;
    \EndFor
    \State
    $\mathcal{NN} \leftarrow \vec{\theta}^{(t_{max})}$;
    \State
    \Return $\mathcal{D_{NN}}$
    \end{algorithmic}
\end{algorithm}

\nnfed{} learns to decode a number-conserving state with only a few randomly generated samples in each iteration. For each update of parameters, the model will get the gradient of loss from the randomly generated dataset. Via gradient descent over the cross-entropy, it learns to decode the whole number-conserving states. This workflow bridges the original scalability guarantee of FED and enables the PQC to be efficiently updated compared to the measurement methods that lacks a decoder. To benchmark the trained \nnfed{}'s performance, we compare its decoding time for a single measurement shot to the genetic algorithm and simulated annealing. We also compare their average success rate over 1000 shots. The results are shown as Table \ref{t1}, where \nnfed{} surpasses both classical algorithms in terms of decoding time and accuracy on the scaling test cases.
\begin{figure*}[bt!]
    \centering
    \hspace*{-0.5cm}\includegraphics[width=0.95\linewidth]{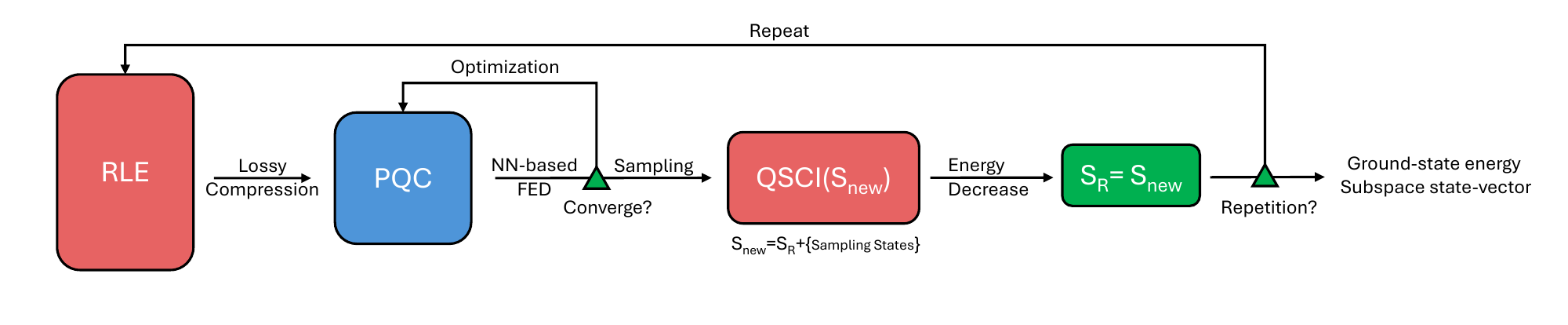}
    \caption{This figure illustrates the \lossyqsci{} workflow. RLE compression is initialized for PQC optimization. The resulting minimum is sampled for contributing configurations. This set of configurations are embedded in the new candidate set $S_{\text{new}}$ if the minimum shows better convergence to the previous optimization trial. }
    \label{workflow}
\end{figure*}

\subsection{Revisit the design of QSCI}

Although both (i) compressing the register below \(\log_{2}\!\bigl[\binom{M}{N}\bigr]\) and (ii) relaxing injectivity checks inevitably discard certain basis states, a carefully designed \chemrle{} eliminates primarily non-physical configurations that do not contribute to the electronic Hamiltonian. If the surviving subspace still contains all determinants that dominate the ground state, the associated accuracy loss is negligible. Assuming this ground state is indeed sparse---an assumption supported by Fig.~\ref{fig1}---we embed \chemrle{} into the QSCI workflow to construct \lossyqsci{}.  

The resulting algorithm (Fig.~\ref{workflow}) introduces an encode--decode layer that (i) samples configurations in the compressed Hilbert space, (ii) decodes them on the fly, and (iii) iteratively updates the CI subspace used for classical diagonalization. This strategy simultaneously reduces the qubit count and enhances sampling efficiency, all while retaining chemical accuracy.

In the \lossyqsci{} workflow, we iteratively refine a compact CI subspace until the ground‐state energy converges (Algorithm~\ref{alg:lossy_qsci}). Each outer iteration comprises five steps:

\smallskip
\noindent\textbf{Compression} - Generate a random \chemrle{} that maps the original \(M\)-orbital register to a \(Q\)-qubit subspace, thereby significantly reducing qubit count.

\smallskip
\noindent\textbf{Decoder training} – Fit a problem-specific \nnfed{} that inverts the new encoding and enables fast evaluation of observables in the full Fock space.

\smallskip
\noindent\textbf{State preparation} – Use a hybrid routine (e.g.\ VQE) to approximate the ground state for energy evaluations with the compressed Hamiltonian and the trained \nnfed{}.

\smallskip
\noindent\textbf{Sampling and decoding} – Draw the \(R\) most probable bit-strings from the prepared state and decode them to \(N\)-electron configurations in the original basis.

\smallskip
\noindent\textbf{Collection} – Augment the candidate set \(S_R\) with the newly decoded configurations to form \(S_{\mathrm{new}}\); diagonalise \(\hat H\) in this enlarged subspace.  If the resulting energy is lower, set \(S_R \leftarrow S_{\mathrm{new}}\) and repeat.

\begin{algorithm}
\caption{\lossyqsci{} Workflow}
\label{alg:lossy_qsci}
\begin{algorithmic}[1]
\Require Hamiltonian \(H\), initial state \(\ket{\Psi}\)
\Ensure Ground-state energy \(E_{\mathrm{best}}\) and configuration subspace \(S_R\)
\Repeat
\State Generate lossy RLE \(E(\cdot)\) to obtain compressed Hamiltonian \(H_{\mathrm{comp}}\)
\State Prepare trial wavefunction \(\ket{\Psi}\) using a PQC under \(H_{\mathrm{comp}}\)
\State Train \nnfed{} to decode compressed bitstrings to \(N\)-electron configurations
\Repeat
  \State Sample bitstrings from \(\ket{\Psi}\)
  \State Decode samples using \nnfed{} and get Energy
  \State Update PQC parameters if required
\Until{convergence criteria met}
\State Select R-most frequent bitstrings to extend new subspace \(S_{\mathrm{new}}\)
\State Diagonalize \(H\) in \(S_R \gets S_R + S_{\mathrm{new}}\) to get energy \(E_{\mathrm{new}}\)
\If{\(E_{\mathrm{new}} < E_{\mathrm{best}}\)}
 \State \(S_R \gets  S_R + S_{\mathrm{new}}\)
 \State \(E_{\mathrm{best}} \gets E_{\mathrm{new}}\)
\EndIf
\Until{convergence criteria met}
\Return \(E_{\mathrm{best}},\, S\)
\end{algorithmic}
\end{algorithm}

The iterative “collect-and-refine’’ loop compensates for the information discarded by the lossy \chemrle{} compression.  After a sufficient number of cycles, all accepted configurations are merged into a final candidate set; the Hamiltonian is then diagonalised classically in this subspace to produce our best ground-state estimate.  By fusing qubit-efficient encoding with targeted quantum sampling and lightweight classical post-processing, the \lossyqsci{} workflow offers a resource-efficient path to chemically accurate energies on today’s noise-limited quantum hardware.

In summary, lossy-QSCI incorporates QSCI with chemically aware lossy-RLE to enhance the sample efficiency and NN-FED to provide efficient expectation measurement. As pointed out by \citet{reinholdt_fundamental_2025}, (i)~sub-space methods such as QSCI require an exponential number of samples if the trial state has significant weight outside the optimal configuration manifold, which further results in (ii)~the classical diagonalization cost will grow prohibitively when the number of redundant selected determinants increases. Lossy-QSCI provides mitigation for both concerns by: First, the chemistry-inspired \chemrle{} concentrates the support of the trial state in an $O(N\log M)$-qubit register; empirically, the required shots thus drop compared with uncompressed QSCI at the same chemical accuracy. Second, by design the lossy map and the $\mathcal{O}(MN)$-parameter NN-FED decoder,it can enable efficient VQE on classical heuristics subspace that can remove redundant selected determinants, so the exact diagonalization (ED) remains a negligible fraction of the total run-time. Hence, the two scalability bottlenecks highlighted in \citet{reinholdt_fundamental_2025} can thus be mitigated when a symmetry-aware lossy encoding and a lightweight decoder are combined with QSCI.

\section{Numerical Result of \lossyqsci{}}
Here we test the lossy compression QSCI on different molecules with the active space selections labeled as \textbf{($N_{\text{Spin Orbitals}}$, $N_{\text{Electrons}}$)} to benchmark the performance of \lossyqsci{}, where $N_{\text{Spin Orbitals}}$ is the number of spin orbitals and $N_{\text{Electrons}}$ the number of electrons. {The ground state support for all cases is constructed by iteratively sampling the $R$ most frequent configurations from iteratively prepared approximate ground states(See Appedix for details).} We first examine the $C_2$ molecule with different lossy compression rates represented to study how the number of qubits scales with the energy in \lossyqsci{}. Then, we test the framework on LiH with the Hardware Efficient Ansatz (HEA) and varying QSCI basis size. Finally, we perform a noisy simulation with a bit-flip error channel on the $H_2$ system with both the typical QSCI and \lossyqsci{} frameworks to investigate the quality of convergence in realistic circumstances. 

{\subsection{$C_2$ Molecule}

We begin by classically diagonalising the full \texttt{6-31G} Hamiltonian for $\mathrm{C}_2$ \citep{hehre_selfconsistent_1972} at a bond length of $0.9 \sim 3\,\text{\AA}$. The resulting eigenvector is taken as the exact ground state,
$\ket{\psi_g}$. To emulate the imperfect trial states that a realistic, noise-affected VQE optimization would deliver, we form a surrogate variational state
\[
   \ket{\psi_n}
   \;=\;
   \frac{\ket{\psi_g} + \ket{\psi_u}}
        {\bigl\lVert \ket{\psi_g} + \ket{\psi_u} \bigr\rVert_2},
\]
where $\ket{\psi_u}$ is a random vector drawn from the same $N$-electron, $M$-orbital sector. This construction yields an overlap $\lvert\braket{\psi_n|\psi_g}\rvert^{2} \simeq 0.85$, in line with fidelities reported for converged VQE circuits of comparable size. Hence, $\ket{\psi_n}$ is adopted as the common input state for all QSCI variants discussed below, providing a VQE-level baseline from which the benefits of \lossyqsci{} can be assessed. Lastly, we select the top 50 most probable states (based on their highest occurrence probabilities) to define the compression ratio $R$.

Figure \ref{fig3} illustrates the performance of \lossyqsci{} as a function of the number of qubits. We benchmark our results against reference energies obtained via exact diagonalization (ED) in smaller active spaces, shifted by 1 kcal/mol (0.00159 Hartree) to denote chemical accuracy thresholds: (i) ED in the (16,4) active space, marked as "Chemical Accuracy (16,4)" (solid blue line); and (ii) ED in the (20,4) active space, marked as "Chemical Accuracy (20,4)" (solid grey line). Discrete markers of different colors and shapes denote the \lossyqsci{} results for qubit counts ranging from 10 to 16:  red triangles for 10 qubits, green crosses for 11 qubits, purple squares for 12 qubits, orange pentagon for 15 qubits, and black circles for 16 qubits.

For lower qubit counts (10 and 11), \lossyqsci{} yields energies below the (16,4) chemical accuracy threshold at shorter bond lengths (indicating superior performance in compact regimes), but exceeds it as the bond length increases and the system begins to dissociate. This trend highlights the varying computational complexity across the dissociation curve and underscores the potential benefits of adaptively selecting qubit counts for compression in different molecular configurations. As the qubit count increases to 12, 15, and 16, the \lossyqsci{} energies progressively approach and often surpass the (20,4) chemical accuracy benchmark, particularly at intermediate and longer bond lengths.

Overall, these results demonstrate that \lossyqsci{}, by compressing a larger active space into a reduced qubit representation, can achieve greater accuracy than ED in equivalently sized or even larger active spaces using the same or fewer qubits. This capability positions \lossyqsci{} as a resource-efficient strategy for quantum simulations, enabling closer approximations to exact quantum state configuration interaction (QSCI) results while mitigating constraints on qubit resources and circuit depth.
\begin{figure}[!htbp]
    \centering
    \hspace*{-0.5cm}\includegraphics[width=1.1\linewidth]{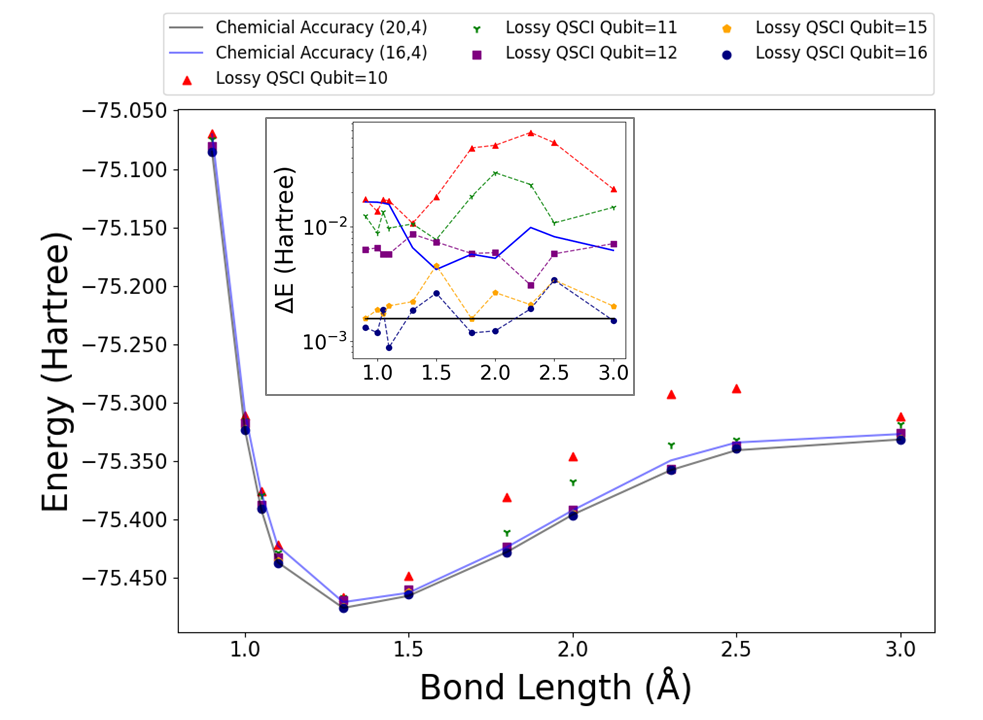}
    \caption{The \( C_2 \) molecule \lossyqsci{} results with approximate ground state prepared by adding random noise to the exact ground state. The x-axis represents the bond length in angstroms, ranging from 0.9 to 3.0, while the y-axis shows the energy in Hartree units, spanning from approximately -75.050 to -75.050. The sub-figure is the energy difference from the exact ground state energy of the active space (20,4).}
    \label{fig3}
\end{figure}}

\subsection{$LiH$ Molecule}

In the study of LiH molecule at a bond length of 2.5 Å with the {STO-3G basis set \citep{hehre_self-consistent_1969}}, the candidate configuration sets, essential for a single round \chemrle{}, is constructed from the 5 most frequently sampled configurations of the approximate ground states prepared via VQE. For the optimization, {we employ L-BFGS-B optimizer\citep{byrd_limited_1995}}.

\begin{figure}
    \centering
    \hspace*{-0.5cm}\includegraphics[width=0.95\linewidth]{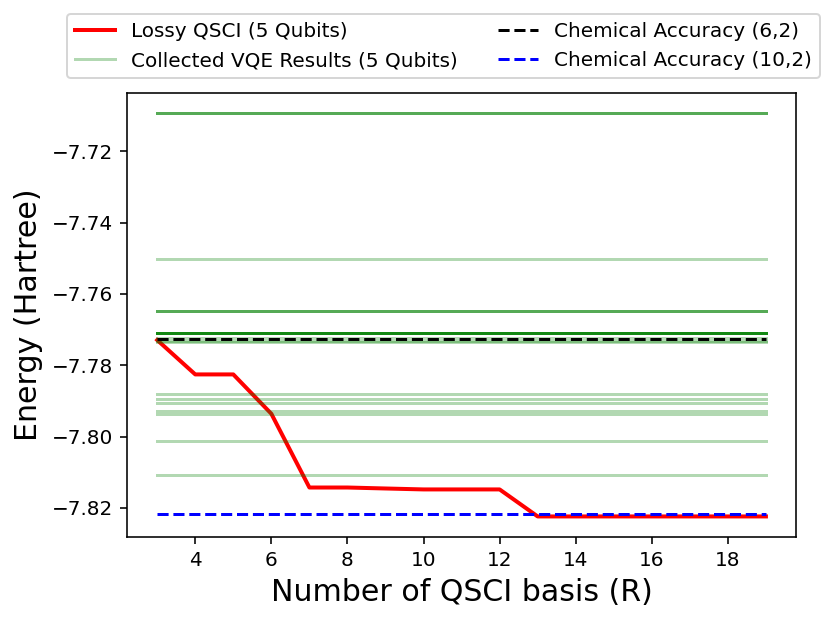}
    \caption{The LiH molecule \lossyqsci{} Results with VQE. The x-axis represents the number of QSCI basis states ($R$) and the y-axis shows the computed energy in Hartree ($\text{Ha}$).}
    \label{fig4}
\end{figure}

Figure \ref{fig4} illustrates the result of the \lossyqsci{} method encoded in five qubits. The two dashed lines, colored in black and blue, provide references for the chemical accuracy in the (6, 2) and (10, 2) active space, respectively. The red line depicts the energy convergence of the \lossyqsci{} method as we increase $R$, starting at slightly above the chemical accuracy of (6,2) active space for $R=4$, converges quickly to the chemical accuracy of (10,2) active space and stabilizes as $R\geq12$. The green line represents the collective optimized energy of VQE on randomly compressed Hamiltonians for five qubits.

The results demonstrate that iterative sampling of CI states from different compressed subspaces (Fig. \ref{workflow}) can significantly enhance the performance of \lossyqsci{}. Specifically, the \lossyqsci{} method achieves the (10,2) chemical accuracy with as few as 12 basis states, outperforming all collective VQE results that lies above -7.8. Improved convergence indicates a potential of \lossyqsci{} to deliver more accurate ground state energies with fewer basis states. This approach achieves the required accuracy with fewer qubit counts compared to conventional QSCI approaches through iterative VQE, thereby improving resource efficiency in quantum simulations, and boosts the simulation power of noisy hardwares.

\subsection{$H_2$ Molecule}
\label{sec:noisy_H2}

We investigate the performance of noisy VQE optimization on the $H_2$ molecule with the 6-31G basis set at bond length $4$Å. We employed a bit-flip error model with single qubit error probability $p_{\text{gate1}} = 0.1$, two-qubit error composed as the tensor product of two single-qubit errors, qubit resetting error -- erroneous preparation of $|1\rangle$ instead of $|0\rangle$ -- $p_{\text{reset}} = 0.1$, and measurements readout error probability $p_{\text{meas}} = 0.1$. The noise model is implemented as Pauli $X$ bit-flip channel via the Qiskit $\texttt{NoiseModel}$ object. Within this setting, we benchmark the \lossyqsci{} algorithm implemented with 4 qubits and a HEA ansatz comprised of 12 CNOT gates and 20 Ry gates against QSCI methods implemented with 8 qubits and a HEA ansatz made up of 14 CNOT gates and 24 Ry gates. 

\begin{figure}
    \centering
    \hspace*{-0.5cm}\includegraphics[width=0.95\linewidth]{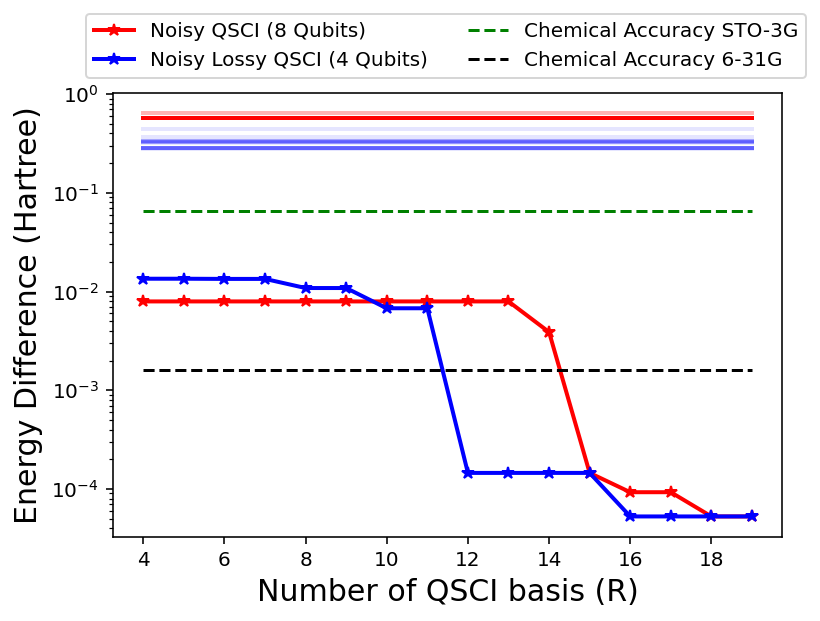}
    \caption{This figure illustrates the $H_2$ molecule noisy simulation. A lossy compressed QSCI simulation not only significantly reduce the qubit cost, but also enhances the convergence w.r.t. increasing QSCI basis. }
    \label{fig5}
\end{figure}

Figure \ref{fig5} illustrates the performance of the \lossyqsci{} and QSCI under the influence of bit-flip error. For the \lossyqsci{} method, the candidate configuration set was constructed by sampling the 10 most frequent configurations from 20 subspaces of ground states (due to random compression) prepared by VQE. On the other hand, the QSCI method's candidate configuration set was constructed by sampling the 10 most frequent configurations from 50 subspaces of ground states prepared by VQE. Since \lossyqsci{} with \chemrle{} is only optimized and sampled on the number-conserving subspace, we post-process the QSCI candidate configuration set only containing the number-conserving state to improve the construction of the effective Hamiltonian.

Both results demonstrate that enforcing number-conserving constraints can provide error robustness for both QSCI and noisy QSCI and enable them to recover from noisy VQE results, as depicted by the red and blue lines. Meanwhile, the \lossyqsci{} method (blue) outperforms QSCI (red), achieving energy below chemical accuracy with as few as 12 basis states compared to the 15 basis states in QSCI. The total sampled configurations from \lossyqsci{} are 200 samples of 4-qubit configurations compared to the 500 samples of 8-qubit configurations for QSCI. This approach highlights the potential of \lossyqsci{} to deliver better searching efficiency. Robustness against bit-flip error shows the potential for making it far more suitable than QSCI in near-term devices.

\section{Conclusion}
In this work, we have introduced the \lossyqsci{} framework, enhanced by the \nnfed{} and the \chemrle{}, to address the challenges of qubit-efficient quantum chemistry simulations on near-term quantum devices. By leveraging the number-conserving properties of fermionic encodings, our framework achieves significant qubit reduction scaling as $O(N \log M)$ for $M$ spin orbitals and $N$ electrons, while mitigating the measurement overhead typically associated with compressed Hamiltonians. Furthermore, the comparison of RLE strategies for the $C_2H_4$ system underscores the advantage of biased chemical encoding, which leverages prior knowledge to improve compression quality. This approach not only allows qubit compression below the theoretical lower bound of $\log_2\binom{M}{N}$ but also minimizes the computational overhead of qubit encoding generation by focusing only on chemically relevant configurations. Inspired by MO Theory and the CAS method, our lossy encoding ensures that critical ground state information is preserved, making it well-suited for quantum chemistry simulations. On the other hand, the integration of NN decoder, as detailed in Algorithm 1, enables fast high precision bitstring decoding of compressed quantum states as shown in Table \ref{t1}, outperforming traditional classical decoders like genetic algorithms and simulated annealing.

Our numerical experiments on the $C_2$, LiH, and $H_2$ molecules demonstrate the efficacy of \lossyqsci{}. {For the $C_2$ molecule, Figure \ref{fig3} shows that \lossyqsci{} achieves energies closer to the exact diagonalization energy curve results of the (20,4) active space as the number of qubits increases from 10 to 16, where 16 qubits results surpass the exact diagonalization benchmark for the (16,4) active space throughout the length of the bond. This highlights the ability of \lossyqsci{} to map larger active spaces to smaller qubit representations without greatly sacrificing accuracy.} Similarly, the LiH molecule illustrates that \lossyqsci{} with five qubits reaches the chemical accuracy of the (10,2) active space with as few as 12 basis states, outperforming the best collected VQE result. The iterative CI states from compressed subspaces, combined with statistical ranking, enable this enhanced performance, as detailed in the improved workflow. Finally, the $H_2$ molecule case shows that the number-conserving subspace prepared by \chemrle{} naturally protects the sample of non-physical state and therefore provides error robustness compared to the original QSCI with number-conserving post-processing.

Overall, our framework demonstrates that \lossyqsci{} can deliver accurate ground state energies with fewer basis states and highlights the potential of near-term devices with fewer qubits and circuit depth to achieve the same accuracy as larger devices through iterative VQE on different random subspaces with classical post-processing. This balance between computational efficiency and accuracy paves the way for resource-efficient ground state simulations on near-term hardware, offering a promising pathway to explore quantum advantages in solving complex many-body problems. While this work demonstrates an improvement of time and space resources for QSCI simulation via efficient encoding-decoding to a smaller subspace, it remains open to characterize the qubit-to-gate trade-offs in the encoded subspace, which will allow us to encode arbitrary QSCI algorithms that leverage both the gate advantages from the unencoded space and subspace simulation. {Beyond it, further research is required to devise provably scalable, symmetry-aware encoders and replace the data-driven decoder algorithms with formal performance bounds. In this work, we present \lossyqsci{} as an intermediate step that enhances near-term resource usage and informs the design of future, fault-tolerant workflows rather than a universal, black-box solution. In future work, we will explore the possibility of an encoder-decoder framework that generalizes the framework of Lossy-QSCI.}

\section{Acknowledgments}
\begin{acknowledgments}
The authors thank Drs. Mathias Mikkelsen, Masaya Kohda, Keita Kanno, and Ming-Zhi Chung (QunaSys Inc.) for their insightful discussions and constructive feedback, which substantially improved the manuscript.
\end{acknowledgments}
\bibliographystyle{apsrev4-2}
\bibliography{Foxconn_RLE}% Produces the bibliography via BibTeX.

\newpage
\appendix
\onecolumngrid
\appendix
{\section*{Appendix: Numerical Details}
For all numerical simulations, the ground state supports are constructed by iteratively sampling from iteratively prepared approximate ground states using either VQE or an approximate ground state that aims to mimic the VQE solution. In each sampling, we will go through the same steps for constructing the ground state support. (1) having the sample probabilities of all sampled configurations. (2) picking the R most frequent configurations (R-largest sample probabilities) as candidates of ground state support. (3) Post-selecting the number-conserving states. (4) add configurations to the existing ground states support and do the exact diagonalization to see if the energy will decrease; if not, keep the previous ground state support.

For illustrating the sampling process, in Figure \ref{fig7}, we take the Lossy-QSCI on the H$_2$ molecule as an example. The computational basis consists of 28 bitstring states, each represents a possible configuration in the number-conserving computing basis of the H$_2$ molecule. These states are listed on the x-axis, ordered by their binary representation (from `00000110` to `11000000`).
The red bars represent the support of the exact ground state probability distribution, derived from the squared amplitudes of the wavefunction ($|\psi_i|^2$). To focus on significant contributions, probabilities below $10^{-3}$ are thresholded to zero, and the heights of the red bars are scaled as $3.5 \times \log(p_i \times 2000)$, where $p_i$ are the probabilities for scaling to align with the data frequency scale. This logarithmic scaling enhances visibility of the ground state support.
The colored stacked bars represent the frequency counts of configurations sampled from 20 subspaces, with each subspace contributing 10 samples, resulting in 200 total samples. Each subspace is assigned a unique color from the `viridis` colormap, labeled as `Sample 0` to `Sample 19` in the legend. The counts are computed by mapping each sampled bitstring to its index in the basis, excluding any bitstrings not present in the 28-state basis to ensure number-conserving. The bars for each subspace are stacked over the red bars.

The complete code, including the dataset, basis states, and plotting parameters, is available in a public repository at [\texttt{https://github.com/ALS15204/lossy-qsci}]. This repository includes the exact bitstring data, the ground state wavefunction amplitudes, ensuring full reproducibility. The Lossy-QSCI parameters, such as the subspace definitions and sampling procedure, are detailed in the repository's documentation, aligned with the methodology described in the main text.
\begin{figure*}[!htbp]
    \centering
    \hspace*{-0.5cm}\includegraphics[width=0.95\linewidth]{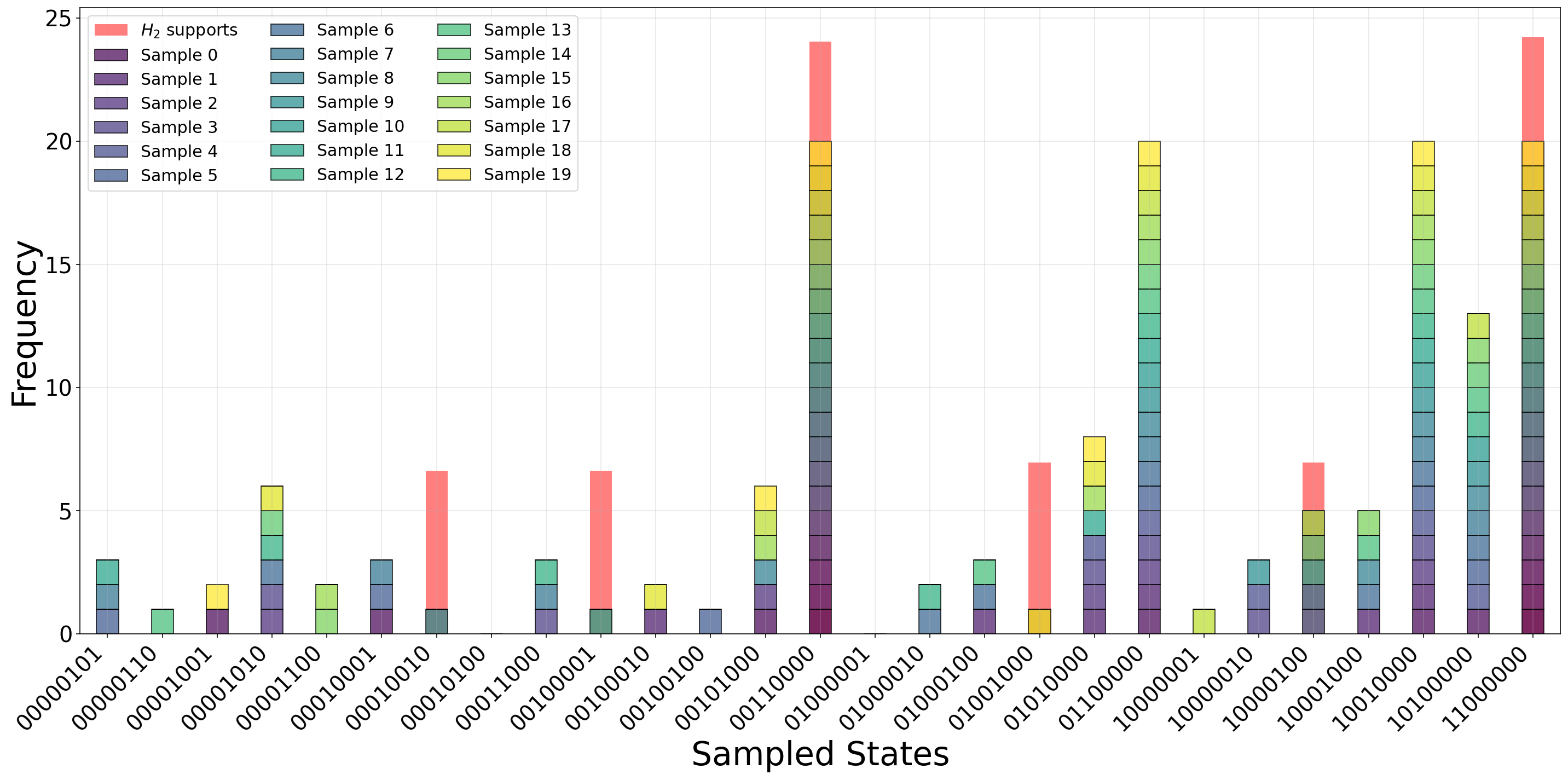}
    \caption{{Configuration distribution for the ground state of the H$_2$ molecule using Lossy-QSCI. Red bars represent the support of the exact ground state probabilities (thresholded at $10^{-3}$, heights proportional to $3.5 \times \log(p_i \times 2000)$). Colored stacked bars show frequency counts of configurations from 20 subspaces (200 total samples, 10 per subspace), with each color indicating a subspace (Sample 0 to 19). The x-axis lists all 28 basis states. The y-axis is the sampling frequency.}}
    \label{fig7}
\end{figure*}}

\end{document}